\documentclass[preprint,showpacs,preprintnumbers,eqsecnum,floatfix]{revtex4}
\usepackage{graphicx}
\usepackage{dcolumn}
\usepackage{bm}

\def\lesssim{\ \raise.3ex\hbox{$<$}\kern-0.8em\lower.7ex\hbox{$\sim$}\ }
\def\gesim{\ \raise.3ex\hbox{$>$}\kern-0.8em\lower.7ex\hbox{$\sim$}\ }
\font\scripti=cmmi7
\font\scriptscripti=cmmi5
\def\sib#1{\setbox0 = \hbox{\scripti #1}
  \kern-.02em\copy0\kern-\wd0
  \kern.04em\box0} 
\def\ssib#1{\setbox0 = \hbox{\scriptscripti #1}
  \kern-.02em\copy0\kern-\wd0
  \kern.04em\box0} 
\font\tenib=cmmib10 
\skewchar\tenib='177 \skewchar\tenib='177 \skewchar\tenib='177
\def\pbold#1{\setbox0 = \hbox{$ #1 $}
  \kern-.022em\copy0\kern-\wd0
  \kern.011em\copy0\kern-\wd0
  \kern.011em\copy0\kern-\wd0
  \kern.011em\copy0\kern-\wd0
  \kern.011em\box0} 
\begin{document}
\title{Single-particle excitations in the BCS-BEC crossover region II: Broad Feshbach resonance}
\author{Y. Ohashi$^{1}$ and A. Griffin$^{2}$}
\affiliation{$^1$Institute of Physics, University of Tsukuba, Tsukuba,
  Ibaraki 305, Japan, \\
$^2$ Department of Physics, University of Toronto, Toronto, Ontario, Canada M5S 1A7}
\date{\today}
\begin{abstract}
We apply the formulation developed in a recent paper [Y. Ohashi and A. Griffin, Phys. Rev. A {\bf 72}, 013601, (2005)] for single-particle excitations in the BCS-BEC crossover to the case of a broad Feshbach resonance. At $T=0$, we solve the Bogoliubov-de Gennes coupled equations taking into account a Bose condensate of bound states (molecules). In the case of a broad resonance, the density profile $n(r)$, as well as the profile of the superfluid order parameter ${\tilde \Delta}(r)$, are spatially spread out to the Thomas-Fermi radius, even in the crossover region. This order parameter ${\tilde \Delta}(r)$ suppresses the effects of low-energy Andreev bound states on the rf-tunneling current. As a result, the peak energy in the rf-spectrum is found to occur at an energy equal to the superfluid order parameter ${\tilde \Delta}(r=0)$ at the center of the trap, in contrast to the case of a narrow resonance, and in agreement with recent measurements. The LDA is found to give a good approximation for the rf-tunneling spectrum. 
\end{abstract}
\maketitle
%

\section{Introduction}
In a recent paper\cite{Ohashi}, we presented a microscopic theory (at $T=0$) of the single-particle excitations in the BCS-BEC crossover regime of a trapped superfluid Fermi gas with a Feshbach resonance. This theory included a Feshbach resonance and the associated molecules. The Bogoliubov single-particle excitations were found by numerically solving the microscopic Bogoliubov-de Gennes (BdG) coupled equations\cite{BdG}. The calculations also involved calculating self-consistently the equation of state given by the Fermi chemical potential (which takes into account the depletion of the number of Fermi atoms as they form into stable molecules). In Ref. \cite{Ohashi}, we applied this theory to a {\it narrow} Feshbach resonance ($g_{\rm r}\sqrt{n}\lesssim \varepsilon_{\rm F}$, where $g_{\rm r}$ is the coupling constant of the Feshbach resonance, $n$ is the number density of atoms, and $\varepsilon_{\rm F}$ is the Fermi energy of a non-interacting Fermi gas). We calculated in a self-consistent manner the density profile, superfluid order parameter, and the single-particle excitation gap $E_g$ in the entire BCS-BEC crossover region. We also discussed the effect of Andreev bound states\cite{Baranov} on the rf-tunneling current data\cite{Chin}.
\par
In this paper, we extend our previous work and present results for single-particle Fermi excitations in the BCS-BEC crossover regime in the case of a {\it broad} Feshbach resonance [$g_{\rm r}\sqrt{n}\gg\varepsilon_{\rm F}$]. This extension is very relevant because all current experiments on superfluid Fermi gases $^{40}$K\cite{Jin} and $^6$Li\cite{Zwierlein,Bartenstein,Kinast,Bourdel} make use of a broad Feshbach resonance in order to tune the pairing interaction. In the case of a broad Feshbach resonance, the BCS-BEC crossover occurs when the bare threshold energy $2\nu$ of the Feshbach resonance is much larger than the Fermi energy $\varepsilon_{\rm F}$, and as a result, the Fermi atoms in the open channel dominate. The Feshbach resonance contributes to an effective pairing interaction between Fermi atoms in the open channel, but the number of stable Feshbach molecules is very small in the crossover region for a broad Feshbach resonance. This situation is quite different from a narrow resonance case, where the Feshbach molecules already dominate in the crossover region. In this paper, we will show how single-particle BCS-like excitations given by the BdG equations differ for the two types of resonances. In particular, we will discuss how this difference shows up in the low-energy spectrum of rf-tunneling current data\cite{Chin}.
\par
We stress that while the differences between the results we find for the excitation spectrum in the case of a broad and narrow resonance are substantial, they emerge from the same microscopic theory based on the Bogoliubov-de Gennes coupled equations. Ultimately the differences arise from quantitative changes in the BdG solutions, such as the spatial width of the order parameter in a trap. 
\par
The numerical solutions of the BdG equations for a broad resonance which we report here are computationally very challenging, because of the need to expand the basis of states to much higher energy. We briefly discuss the associated cutoff problems in Sec. II.
\par
As has been discussed extensively in the recent literature, a broad Feshbach resonance can be treated using the single-channel model, pioneered by Leggett\cite{Leggett}, Nozi\`eres and Schmitt-Rink\cite{Nozieres}, Randeria\cite{Randeria}, and coworkers\cite{Melo,Engelbrecht}. Ref.\cite{Ohashi} and the present paper are built on a coupled fermion-boson two-channel model (see Sec. II). This is necessary to deal with a narrow Feshbach resonance, which was treated in Ref.\cite{Ohashi}. We think that treating a broad Feshbach resonance using a two-channel model, while not necessary, has an advantage in that it allows one to see how the difference between a broad and narrow Feshbach resonance emerge within a unified formalism. However, we note that within a single-channel model, our BdG equation approach can be viewed as an extension of the original work by Bruun\cite{Bruun}, which only considered the BCS limit of the BCS-BEC crossover region.
\par
This paper is organized as follows. Since the formalism presented in Ref. \cite{Ohashi} is valid for both narrow and broad Feshbach resonances, we refer to this earlier paper for most details. In Sec. II, we briefly summarize the coupled fermion-boson model, which our analysis is based upon. The proper treatment of higher energy states in the numerical calculations for the case of a broad Feshbach resonance is noted. In Sec. III, we present our self-consistent numerical solutions of the BdG equations for a broad Feshbach resonance. We discuss single-particle excitations in Sec. IV and use these results to calculate the rf-tunneling current response in Sec. V.
\vskip3mm
\section{Coupled fermion-boson model}
\vskip3mm
\par
A superfluid Fermi gas with a Feshbach resonance can be described by the coupled fermion-boson model\cite{Ohashi,Timmermans,Holland,Ranninger},
\begin{eqnarray}
H
&=&
\sum_\sigma\int d{\bf r}
\Psi^\dagger_\sigma({\bf r})
\Bigl[
-{\nabla^2 \over 2m}-\mu+V_{\rm trap}^{\rm F}({\bf r})
\Bigr]
\Psi_\sigma({\bf r})
-U
\int d{\bf r}
\Psi^\dagger_\uparrow({\bf r})
\Psi^\dagger_\downarrow({\bf r})
\Psi_\downarrow({\bf r})
\Psi_\uparrow({\bf r})
\nonumber
\\
&+&
\int d{\bf r}
\Phi^\dagger({\bf r})
\Bigl[
-{\nabla^2 \over 2M}+2\nu-\mu_M+V_{\rm trap}^{\rm M}({\bf r})
\Bigr]
\Phi({\bf r})
+
g_{\rm r}
\int d{\bf r}
\Bigl[
\Phi^\dagger({\bf r})\Psi_\downarrow({\bf r})\Psi_\uparrow({\bf r})+h.c.
\Bigr].
\nonumber
\\
\label{eq.2.1}
\end{eqnarray}
We refer to Ref.\cite{Ohashi} for further discussion about this model.
We assume a two-component Fermi gas described by a fermion field operator $\Psi_\sigma({\bf r})$, where the pseudo-spin label $\sigma=\uparrow,\downarrow$ represent two atomic hyperfine states. These states are in the so-called open-channel. $\Phi({\bf r})$ describes the molecular bosons associated with a Feshbach resonance. $g_{\rm r}$ is a coupling constant describing the Feshbach resonance, and $2\nu$ is the bare threshold energy of the resonance. Experimentally, $2\nu$ is tunable by an external magnetic field, so that the effective pairing associated with the Feshbach resonance is also tunable. This interaction becomes stronger as we decrease the threshold energy $2\nu$. The atoms and molecules are trapped in harmonic potentials $V_{\rm trap}^{\rm F}({\bf r})\equiv (1/2)m\omega_0^2r^2$ and $V_{\rm trap}^{\rm M}({\bf r})\equiv (1/2)M\omega_0^2r^2$, respectively. Here, we assume an isotropic trap, for simplicity, and assume that atoms and molecules feel the same trap frequency $\omega_0$ (as is the case in current experiments). 
\par
Since a molecule consists of two Fermi atoms, the molecular mass is twice as large as a atomic mass, $M=2m$. We also impose the conservation of the total number of atoms,  
\begin{eqnarray}
N
=
N_{\rm F}+2N_{\rm M}
=
\int d{\bf r} n_{\rm F}(r)
+
2\int d{\bf r} n_{\rm M}(r),
\label{eq.2.2}
\end{eqnarray}
where $N_{\rm F}$ and $N_{\rm M}$ represent the number of Fermi atoms in the open channel and the number of molecules associated with the Feshbach resonance, respectively. Equation (\ref{eq.2.2}) defines the number density of Fermi atoms in the open channel $n_{\rm F}(r)$ and the local density $n_{\rm M}(r)$ of molecules. Since we limit ourselves to zero temperature in this paper, all the molecules are Bose-condensed. The constraint in Eq. (\ref{eq.2.2}) has already been taken into account in Eq. (\ref{eq.2.1}) by taking the Bose chemical potential $\mu_M\equiv 2\mu$, where $\mu$ is the Fermi chemical potential. Equation (\ref{eq.2.1}) also includes a non-resonance interaction $U$, which is taken to be weakly attractive.
\par
The BCS-BEC crossover at $T=0$ is treated in Ref.\cite{Ohashi} by solving the BdG coupled equations for the mean-field approximation to the Hamiltonian in Eq. (\ref{eq.2.1}), together with the equation for the number of atoms\cite{Leggett}. Using Eq. (\ref{eq.2.1}), the superfluid phase is characterized by the composite order parameter\cite{Ohashi,Ranninger,Timmermans,Holland,Ohashi2},
\begin{eqnarray}
{\tilde \Delta}(r)\equiv\Delta(r)-g_{\rm r}\phi_M(r).
\label{eq.2.3}
\end{eqnarray}
Here, $\Delta(r)\equiv U\langle\Psi_\downarrow({\bf r})\Psi_\uparrow({\bf r})\rangle$ is the Cooper-pair order parameter, and $\phi_M\equiv\langle\Phi({\bf r})\rangle$ describes the molecular condensate. For details about the self-consistent calculation of ${\tilde \Delta}(r)$, we refer to Ref.\cite{Ohashi}.
\par
In this paper, we take ${\bar g}_{\rm r}\equiv g_{\rm r}\sqrt{N/R_{\rm F}^3}=10\varepsilon_{\rm F}$ to describe a broad Feshbach resonance, where $R_{\rm F}\equiv\sqrt{2\varepsilon_{\rm F}/m\omega_0^2}$ is the Thomas-Fermi radius. This is the only parameter change from Ref.\cite{Ohashi}, which considered a narrow resonance with ${\bar g}_{\rm r}=0.2\varepsilon_{\rm F}$.
The total number of atoms is again taken to be $N=10912$, which corresponds to the Fermi energy $\varepsilon_{\rm F}=31.5\omega_0$, where $\omega_0$ is the trap frequency. For the non-resonant interaction $U$, we take $U(N/R_{\rm F}^3)=0.35\varepsilon_{\rm F}$. 
\par
In solving the BdG equations numerically, one cannot retain all the harmonic potential eigenfunctions. Thus we need to truncate the basis states by introducing a finite cutoff energy $\omega_c$. In this paper, we take $\omega_c=161.5\omega_0=5.1\varepsilon_{\rm F}$. To avoid any spurious effects of this cutoff, we need to restrict our calculations to the region where the maximum value of the composite order parameter ${\tilde \Delta}(r)$ in the center of the trap is {\it much smaller} than this cutoff $\omega_c$. In the case of a narrow Feshbach resonance ($g_{\rm r}\sqrt{N/R_{\rm F}^3}\lesssim \varepsilon_{\rm F}$), ${\tilde \Delta}(r)$ does not become very large even in the BEC regime, so that this computational restriction is not a serious problem. However, ${\tilde \Delta}(r)$ in a broad Feshbach resonance ($g_{\rm r}\sqrt{N/R_{\rm F}^3}\gg\varepsilon_{\rm F}$) can be very large in the BEC regime. Thus, in this paper, we restrict our calculations in the strong-coupling BEC regime to the region $(k_{\rm F}a_s)^{-1}\lesssim 1$, where ${\tilde \Delta}(r=0)\sim O(\varepsilon_{\rm F})$ (here $k_{\rm F}$ is the Fermi momentum, and $a_s$ is the two-body atomic $s$-wave scattering length). In the equation of state involving the number of atoms, we still take into account the contribution coming from the high energy states with energies $\omega>\omega_c$ by using a local density approximation (LDA). These high energy contributions are important, since neglecting them means that the number of molecules $n_{\rm M}$ is overestimated in the BEC regime.
\par
\section{Density profile and composite order parameter}
\par
Figure 1 shows our results for the Fermi chemical potential $\mu$ in the BCS-BEC crossover. This is very similar to the result for a narrow Feshbach resonance (see Fig. 6 in Ref. \cite{Ohashi}). That is, for both broad and narrow Feshbach resonances, the chemical potential $\mu$ has qualitatively the same dependence on the value of the parameter $(k_{\rm F}a_s)^{-1}$, where $a_s$ is the two-body scattering length. We note that $\mu$ gradually deviates from being equal to the Fermi energy $\varepsilon_{\rm F}=31.5\omega_0$ as we increase the pairing interaction, and changes sign at $(k_{\rm F}a_s)^{-1}\simeq 0.65$. For a broad Feshbach resonance, this change is found to correspond to $\nu\simeq 50\varepsilon_{\rm F}$ (see the inset of Fig. 1). This threshold energy is much higher than in the case of a narrow resonance (see Fig. 6 of Ref.\cite{Ohashi}), where we found that $\mu=0$ at $2\nu\simeq 0$.
For a broad resonance, we might note that the generalized $s$-wave scattering length $a_s$ defined in Eq. (4.5) of Ref.\cite{Ohashi} reduces to the usual two-body scattering length $a_s^{2b}$ for a Feshbach resonance defined in Eq. (4.7) of Ref.\cite{Ohashi}. This is because we have $2\nu\gg2\mu$ in the crossover region for a broad resonance.
\par
Figure 2 shows the total atomic density profile $n(r)=n_{\rm F}(r)+2n_{\rm M}(r)$. In the BCS regime, the density profile is significant right up to the Thomas-Fermi radium $R_{\rm F}$, which describes the size of a non-interacting Fermi gas in a harmonic trap at $T=0$. The density profile gradually shrinks as one approaches the (strong-coupling) BEC regime. However, this narrowing is much less than we found in the case of a narrow resonance (see Fig. 5 in Ref. \cite{Ohashi}). In a broad resonance, the Fermi atoms in the open channel are still the dominant particles in the crossover region. (As noted above, this occurs at large values of $2\nu$, as shown by the inset in Fig. 1.) As a result, repulsion resulting from the Pauli exclusion principle between Fermi atoms gives a wider atomic density profile in the case of a broad Feshbach resonance. This is quite different in the case of a narrow resonance, where the Feshbach molecules (bosons) already dominate in the crossover region, and the Pauli principle is less effective. As a result, the spatial profile of $n(r)$ is more localized at the center of the trap in a narrow resonance.
\par
Although the Fermi atoms in the open channel are dominant in the broad Feshbach resonance, we note that even at $T=0$, only a small fraction of the atoms are associated with Cooper-pairs. Figure 3 shows the Cooper-pair condensate fraction $N_{\rm C}^F$, defined by\cite{Parola,Gio}
\begin{eqnarray}
N_{\rm C}^F\equiv\int d{\bf r}d{\bf r}'|\langle\Psi_\downarrow({\bf r})\Psi_\uparrow({\bf r}')\rangle|^2.
\label{eq.2.4}
\end{eqnarray}
In the BCS regime, Fig. 3 shows that only a small fraction of $N_{\rm F}$ contributes to the Cooper-pair condensate fraction $N_{\rm C}^F$. This reflects the fact that only atoms near the Fermi surface form a Cooper-pair Bose-condensate in the weak-coupling BCS regime. Atomic states well below the Fermi energy are essentially unmodified from those in a non-interacting Fermi gas. The Cooper-pair condensate fraction $N_{\rm C}^F$ starts to increase as one approaches the BEC regime. At $(k_{\rm F}a_s)^{-1}\simeq 1$, we find $2N_{\rm C}^F\simeq N_{\rm F}$, indicating that most atoms in the open channel form Cooper-pairs which are Bose-condensed. Namely, this regime can be regarded as the BEC of bound pairs of atoms from the open channel. 
\par
Figure 4 shows the spatial variation of the composite order parameter ${\tilde \Delta}(r)$ in the crossover region. Although the number of Feshbach molecules $N_{\rm M}$ is very small (see Fig. 3), we find that the molecular BEC order parameter $-g_{\rm r}\phi_M(r)$ is comparable to the Cooper-pair component $\Delta(r)$. Since the coupling $g_{\rm r}$ is large in a broad Feshbach resonance, the contribution of the molecular order parameter $-g_{\rm r}\phi_M(r)$ is clearly enhanced.
Even in the BCS region at $(k_{\rm F}a_s)^{-1}=-1$ [where $N_{\rm M}/N_{\rm F}\sim 10^{-6}$, as in Fig.3], Fig 4(a) shows that $-g_{\rm r}\Phi_M(0)/{\tilde \Delta}(0)\simeq 0.24$. In Fig. 4(d), the molecular BEC component becomes dominant over the Cooper-pair component, although the number of condensed Feshbach molecules is still very small, as shown in Figs. 2(d) and 3. Thus, even in the case of a broad Feshbach resonance, the Feshbach molecules are found to make a major contribution to the composite order parameter ${\tilde \Delta}(r)$, although the number of the Feshbach molecules is very small.
This surprising result is due to the fact that the Cooper-pair order parameter $\Delta(r)=\langle\Psi_\downarrow(r)\Psi_\uparrow(r)\rangle$ is quite different from the number density of atoms $n_{\rm F}(r)=\sum_\sigma\langle\Psi_\sigma^\dagger(r)\Psi_\sigma(r)\rangle$ in a Fermi superfluid. The situation is quite different from a Bose gas BEC, where there is a simple relation between the BEC order parameter $\phi_M(r)=\langle\Phi(r)\rangle$ and the number density of condensed bosons at $T=0$, namely $n_M(r)=\langle\Phi^\dagger(r)\Phi(r)\rangle=|\phi_M(r)|^2$. The fact that $n_{\rm M}(r)/n_{\rm F}(r)\ll 1$ does not necessarily mean a small value of the ratio $-g_{\rm r}\phi_M(r)/\Delta(r)$. 
\par
\section{Single-particle excitation spectrum}
\par
Figure 5 shows the fermion single-particle local density of states, defined by\cite{Ohashi}
\begin{eqnarray}
N(\omega,r)=-{1 \over \pi}{\rm Im}[G_{11}({\bf r},{\bf r},i\omega_n\to\omega+i\delta)].
\label{eq.3.1}
\end{eqnarray}
Here $G_{11}$ is the analytic-continued BCS-like single-particle thermal Green's function,
\begin{equation}
G_{11}({\bf r},{\bf r}',i\omega_n)=-\int_0^\beta d\tau e^{i\omega_n\tau}
\langle T_\tau\{\Psi_\uparrow({\bf r},\tau)\Psi_\uparrow^\dagger({\bf r}',0)\}\rangle,
\label{eq.3.2}
\end{equation}
where $\omega_n$ is the fermion Matsubara frequency. Equation (\ref{eq.3.1}) describes single-particle excitations of energy $\omega$ at position ${\bf r}$.
\par
As discussed in Ref. \cite{Ohashi}, atoms in a trapped Fermi superfluid feel the effect of a combined potential well, consisting of a diagonal trap potential $V_{\rm trap}^F(r)-\mu$ and an off-diagonal pair potential ${\tilde \Delta}(r)$. In the BCS regime, this combined potential well has a minimum at $r\simeq R_{\rm F}$ (where $R_{\rm F}$ is the Thomas-Fermi radius). The Andreev bound states\cite{Ohashi,Baranov} are then formed near the bottom of this potential well, which is near the edge of the trapped gas ($\sim R_{\rm F}$). Indeed, we find some Andreev bound states localized around $r\sim R_{\rm F}$ in Fig. 5(a), and the lowest one gives the single-particle energy gap $E_{\rm g}~(\sim \omega_0)$ of this trapped Fermi superfluid. We note that $E_g$ is much smaller than the typical magnitude of the composite order parameter at the center of the trap. [In Fig. 5(a), we find that ${\tilde \Delta}(r=0)\sim 10\omega_0\gg E_g$.]. 
In Fig. 5(a), we see that there is a large excitation gap ($\sim10\omega_0\gg E_g$) near the center of the trap ($r\sim 0$). This large energy gap in $N(\omega,r)$ is seen to be equal to the magnitude of the composite order parameter ${\tilde \Delta}(r=0)$ in the center of the trap. Thus while the true threshold energy of the single-particle excitation spectrum is determined by the lowest Andreev states at the edge of the trap, the local excitation gap at the center of the trap is equal to the value of the local composite order parameter ${\tilde \Delta}(r=0)$. In the BCS regime, since $E_g$ is largely determined by the width of the combined potential well, $E_g$ is not very sensitive to the increase of the composite order parameter ${\tilde \Delta}(r=0)$ in the center of the trap. As a result, in the BCS regime, $E_g$ only slowly increases as we go from the BCS region to the crossover regime (where $(k_{\rm F}a_s)^{-1}\sim 0$). This is to be compared with the rapid increase of ${\tilde \Delta}(r=0)$ on the BEC side, as shown in Fig. 6.
\par
In the BEC regime, when the Fermi chemical potential $\mu$ is negative,  Fig. 5(b) shows that the local density of states $N(\omega,r)$ has a large energy gap everywhere, including at the edge of the trap. The physics of this can be easily understood, using the Bogoliubov single-particle excitation spectrum for a {\it uniform} Fermi superfluid given by
\begin{equation}
E_{\bf p}=\sqrt{(p^2/2m-\mu)^2+{\tilde \Delta}^2}.
\label{eq.3.3}
\end{equation}
When $\mu<0$, this energy spectrum $E_{\bf p}$ has an energy gap\cite{Leggett} given by $E_g=\sqrt{\mu^2+{\tilde \Delta}^2}$. In the (strong-coupling) BEC limit, this gap approaches $|\mu|$. Namely, instead of the superfluid order parameter ${\tilde \Delta}$, it is the magnitude of the Fermi chemical potential which determines the single-particle excitation gap. In turn, this determines the binding energy of a Cooper pair. Similarly, in a trapped gas, even though the order parameter is small near the edge of the gas, $E_g$ is still large in the BEC regime due to a large magnitude of the chemical potential $|\mu|$. Figure 6 shows how the calculated energy gap $E_g$ approaches $|\mu|$ in the BEC regime ($\mu<0$), as expected.
\par
In a narrow Feshbach resonance considered in Ref.\cite{Ohashi}, the single-particle excitation gap $E_g$ has a minimum value when one is in the crossover region (see Fig. 11 of Ref.\cite{Ohashi}). This is due to the significant narrowing of the order parameter spatial profile in the crossover region. For a {\it narrow} resonance, the width of the combined potential well becomes wider in the crossover region, and consequently the energy of the lowest Andreev state decreases. On the other hand, for a {\it broad} Feshbach resonance, the narrowing of the order parameter spatial profile is not significant (see Fig. 4). That is, the width of the combined potential well does not become wider, in contrast with the case of a narrow resonance. As a result, the single-particle energy gap $E_g$, determined by the lowest Andreev state, steadily increases as we move from the BCS to BEC region. However, this increase of $E_g$ is still slow in the BCS regime, compared with the rapid increase in ${\tilde \Delta}(r=0)$, as shown in Fig. 6. In the next section, we will point out that this quite different behavior of $E_g$ in the case of a broad and narrow Feshbach resonance has important implications for the rf-tunneling current spectroscopy.
\par
\section{rf-tunneling current spectrum}
\par
In Figs. 7 and 8, the solid line shows the calculated rf-tunneling current spectrum $I_F(\omega)$ in the BCS-BEC crossover region. Here $\omega\equiv\omega_L-\omega_a-\mu+\mu_a$ is the effective detuning frequency. The frequency of the laser light is $\omega_L$, while $\omega_a$ and $\mu_a$ are the threshold energy and chemical potential, respectively, of the additional atomic hyperfine state $|a\rangle$ which the open state $\downarrow$ is coupled to.
For the detailed derivation and discussion of the expression for $I_F(\omega)$ we use, based on the solution of the microscopic BdG equations, see Sec. VII of Ref. \cite{Ohashi}.
\par
In a broad Feshbach resonance, we find that the peak energy in the rf-spectrum increases with increasing pairing interaction [panels (a)-(c)]. In the BEC regime [panel (d)], the spectrum $I_{\rm F}(\omega)$ also exhibits a finite energy gap. This energy gap reflects the large single-particle excitation gap in the BEC regime, shown in Fig. 5(b) and discussed in Sec. IV.
\par
In Fig. 6, the solid circles show the peak energies of the {\it calculated} rf-tunneling current spectra in the crossover region (as shown by the examples in Figs. 7 and 8). These $T=0$ results clearly indicate that for a broad resonance, the peak energy in the rf-spectrum can give a direct measurement of the magnitude of the order parameter ${\tilde \Delta}(r=0)$ in the center of the trap. Figure 6 also shows the {\it observed} peak energies in the recent rf-tunneling experiment on superfluid $^6$Li, as open circles\cite{Chin}. Although the present calculation is for an isotropic trap at $T=0$ while the experiments have been done in a cigar-shaped trap at finite temperatures, we note that our results agree reasonably well with the experimental data, and especially well in the BEC region [i.e., in the region $(k_{\rm F}a_s)^{-1}\gesim 0.5$].
\par
In the case of a narrow Feshbach resonance\cite{Ohashi}, the very low-energy rf-spectrum in the BCS regime is dominated by Andreev excitations localized near the edge of the trap. As discussed in detail in Sec. VII of Ref.\cite{Ohashi}, the result is that the contribution of the large order parameter in the center of the trap is hidden in $I_{\rm F}(\omega)$ by the spectral weight of the low-energy surface excitations. 
This effect is still significant in the crossover region. 
This is because the spatial profile of the composite order parameter ${\tilde \Delta}(r)$ narrows in the crossover region, as shown in Fig. 5 of Ref. \cite{Ohashi}. As we have discussed in Sec. IV above, this narrowing lowers the energies of the Andreev bound states localized at the bottom of the combined potential well. Thus for a narrow resonance, these low-energy Andreev states dominate the quasi-particle excitation spectrum, with the result that the low-frequency spectrum of $I_{\rm F}(\omega)$ is also dominated by these surface excitations. In the BEC regime (where $\mu<0$), we recall that the effects of these low-energy Andreev-like excitations are suppressed. In a {\it broad} Feshbach resonance, in which the spatial narrowing of the profile of ${\tilde \Delta}(r)$ was found to be not significant, the role of the low-energy quasi-particle excitations at the edge of the trap is less dominant, compared with the narrow resonance case. Our results thus verify\cite{Torma} that for a broad Feshbach resonance, the peak in the rf-tunneling current spectrum gives a direct measurement of the order parameter at the center of the trap. The effect of the low energy Andreev bound states is negligible on $I_{\rm F}(\omega)$.
\par
In principle, one could try to observe the true single-particle energy gap $E_g$ from the threshold energy of the rf-tunneling spectrum. For example, in Fig. 7(b), $E_g$ is given by the lowest peak in the plotted rf-spectrum. Since $E_g$ is very small, high resolution would be necessary to observe this small gap structure in the rf-tunneling spectrum. However, we stress that the these Andreev surface states are of great interest, representing as they do a very interesting aspect of the quasiparticle spectrum of a Fermi superfluid.
\par
In Figs. 7 and 8, the dashed line shows our results using for $I_{\rm F}(\omega)$ calculated in the local density approximation (LDA). (The LDA expression for $I_F(\omega)$ is given by Eq. (7.19) in Ref.\cite{Ohashi}.) $I_{\rm F}(\omega)$ was first computed using a LDA in Ref. \cite{Torma}. Our LDA calculation is somewhat different, since self-consistent BdG solutions for ${\tilde \Delta}(r)$ and $\mu$ are used. We find that when the self-consistent solution of the BdG equations are used, the LDA (dashed lines) gives a good overall approximation to the fully microscopic calculations (solid lines) in the crossover region. In particular, the peak energy in the LDA result for $I_{\rm F}(\omega)$, which corresponds to the magnitude of the composite order parameter ${\tilde \Delta}(r=0)$ in the center of the trap, is in excellent agreement with the microscopic result based on the BdG eigenstates. 
\par
\vskip2mm
\section{Summary}
In this paper, we have extended our previous theoretical calculations\cite{Ohashi} for a narrow Feshbach resonance to the case of a broad Feshbach resonance. This extension is important because all current experiments make use of a broad Feshbach resonance. We have numerically solved the BdG coupled equations, together with the equation of state for the number of atoms, in the BCS-BEC crossover region at $T=0$. This formalism we use was developed in detail in Ref.\cite{Ohashi}. The case of a broad resonance involves much more demanding numerical calculations because of the need to include high energy states in solving the BdG equations.
\par
For a broad Feshbach resonance ($g_{\rm r}\sqrt{N/R_{\rm F}^3}\gg\varepsilon_{\rm F}$), the BCS-BEC crossover occurs in the region where the threshold energy $2\nu$ of the Feshbach resonance is much larger than the Fermi energy $\varepsilon_{\rm F}$ (see Fig. 1). As a result, molecules associated with the Feshbach resonance only contribute to an effective paring interaction between atoms in the open channel in a virtual sense. As expected, the number of Feshbach molecules is found to be very small compared to the number of Cooper-pairs. Because of this, the effect of Pauli exclusion principle between Fermi atoms is still important in the crossover region. As a result, the spatial profile of the composite order parameter ${\tilde \Delta}(r)$ is spread out and broad even in the crossover region. This is quite different from the narrow resonance case\cite{Ohashi}, where Feshbach molecules dominate in the crossover region. In this case, ${\tilde \Delta}(r)$ is spatially more centered in the trap.
\par
This difference in the width of the spatial profile of the composite order parameter ${\tilde \Delta}(r)$ between a broad and narrow Feshbach resonances leads to qualitatively different results for the low-energy rf-tunneling current spectrum. For a broad Feshbach resonance, the narrowing of the spatial profile of ${\tilde \Delta}(r)$ is not significant in the crossover region. As a result, the energies of the Andreev bound states localized at the edge of the trap are not lowered and these states end up having little weight in the rf-tunneling current spectrum $I_{\rm F}(\omega)$. As a result, the peak energy in the rf-spectrum is predicted to occur at the value of the composite order parameter ${\tilde \Delta}(r=0)$ at the center of the trap. In contrast, in a narrow Feshbach resonance, the Andreev bound states strongly modify the low-energy rf-tunneling current spectrum in the crossover region. This hides the appearance of the peak at ${\tilde \Delta}(r=0)$ in $I_{\rm F}(\omega)$, as discussed in Ref.\cite{Ohashi}.
\par
In summary, for a broad resonance, we have solved the BdG equations for the coupled fermion-boson model in Eq. (\ref{eq.2.1}) at $T=0$. We have taken into account the formation of a composite order parameter composed of both Cooper-pairs and a BEC of real molecules.
The formalism we use was developed in Ref.\cite{Ohashi}, where it was applied to the case of a narrow Feshbach resonance. In this paper, we used our BdG eigenstates to calculate the rf-tunneling current and found excellent overall agreement with the earlier work based on a simpler LDA\cite{Torma}. In particular, we confirm at a more microscopic level (at $T=0$) that the finite energy peak\cite{Chin} which develops in the rf-tunneling current in the superfluid phase can be used to give a direct measurement of the magnitude of the order parameter at the center of the trap. 
\par
\acknowledgments
Y. O. was financially supported by a Grant-in-Aid for Scientific Research from the Ministry of Education, Culture, Sports, Science and Technology of Japan, as well as by a University of Tsukuba Nanoscience Research Project. A. G. acknowledges a research grant from NSERC of Canada.
\par 
%
%
\newpage


%
\newpage
\begin{figure}
\includegraphics[width=10cm,height=8cm]{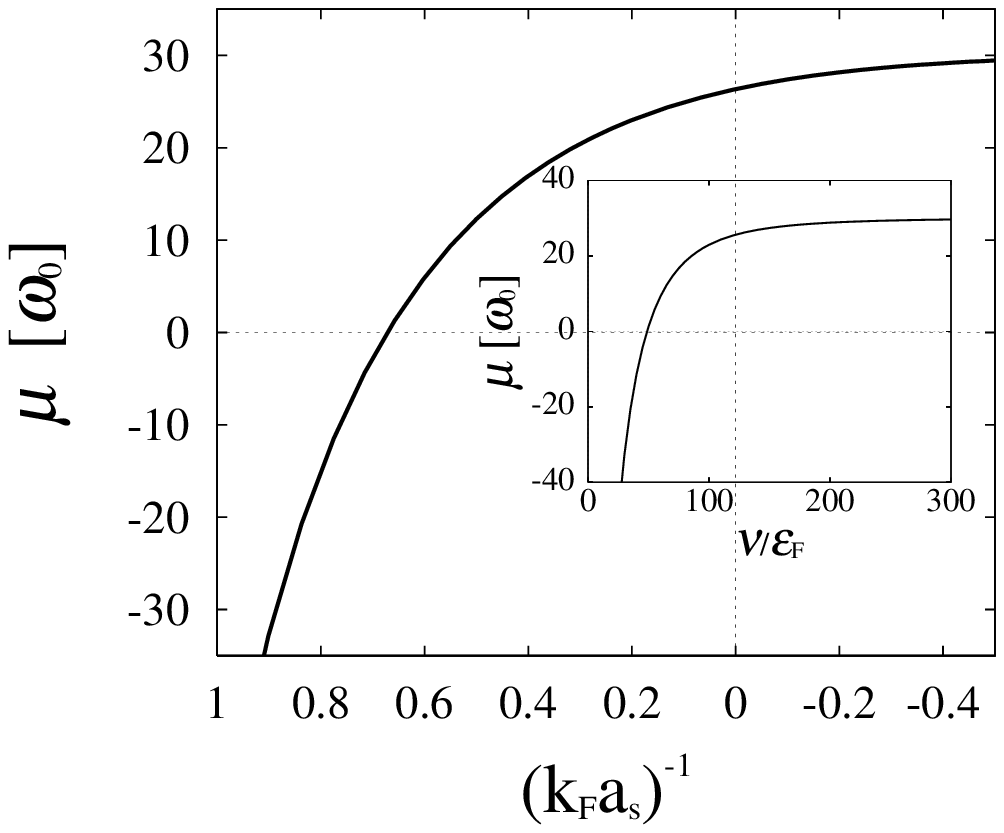}%
\caption{
Fermi chemical potential $\mu$ in the BCS-BEC crossover region for a broad Feshbach resonance ($g_{\rm r}\sqrt{N/R_{\rm F}^3}=10\varepsilon_{\rm F}$). The interaction is given in terms of the two-body $s$-wave atomic scattering length $a_s$\cite{Ohashi}. The inset shows the same chemical potential $\mu$ plotted as a function of the bare threshold energy $2\nu$ of the Feshbach resonance.
\label{fig1} 
}
\end{figure}

\begin{figure}
\includegraphics[width=16cm,height=16cm]{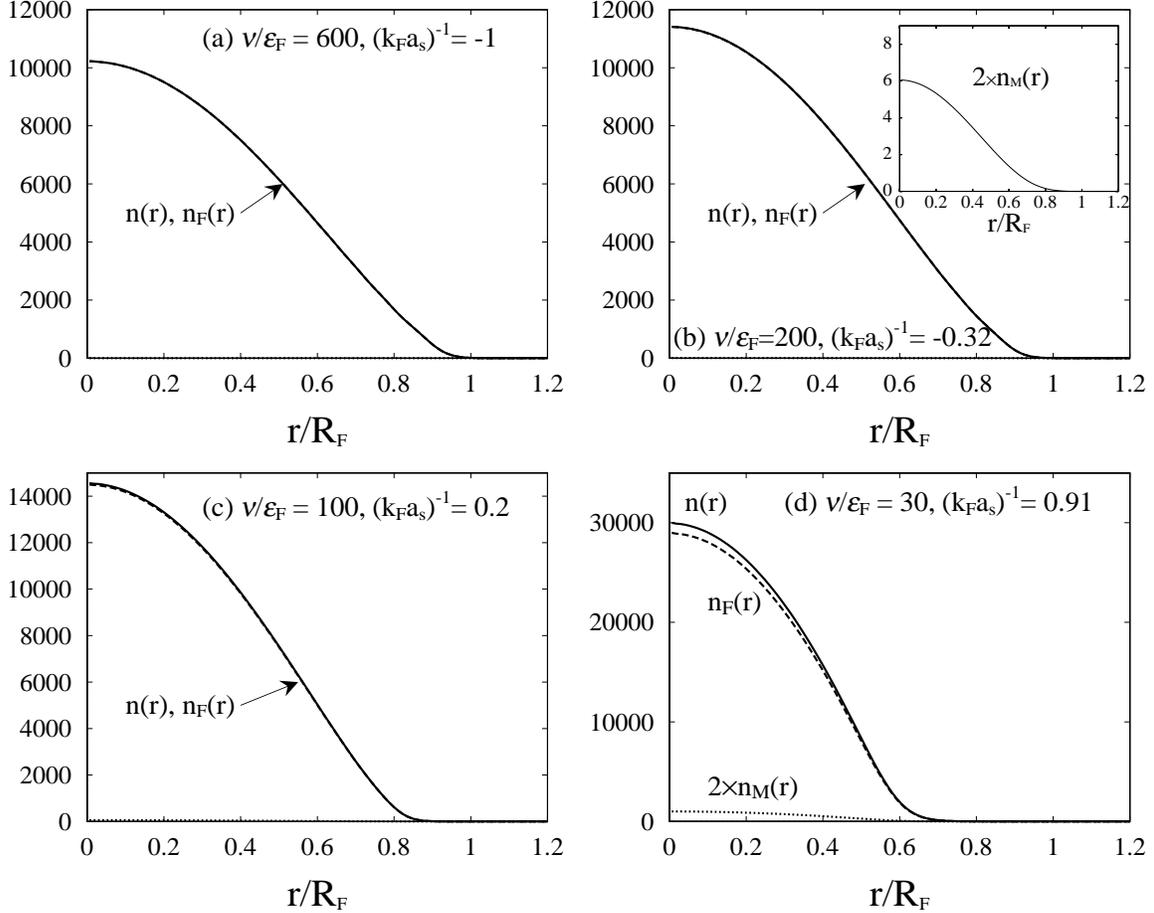}%
\caption{
Total atomic density profile $n(r)$ in the BCS-BEC crossover for a broad resonance. We also show the atomic density profile in the open channel $n_{\rm F}(r)\equiv\sum_\sigma\langle\Psi_\sigma^\dagger({\bf r})\Psi_\sigma({\bf r})\rangle$, as well as the molecular condensate density associated with the Feshbach resonance $n_M(r)\equiv|\phi_M({\bf r})|^2$. The molecular condensate is very small in panels (a)-(c) [$n_M({\bf r})\ll n(r)]$. The inset in panel (b) shows the molecular density profile in more detail.
\label{fig2} 
}
\end{figure}

\begin{figure}
\includegraphics[width=10cm,height=13cm]{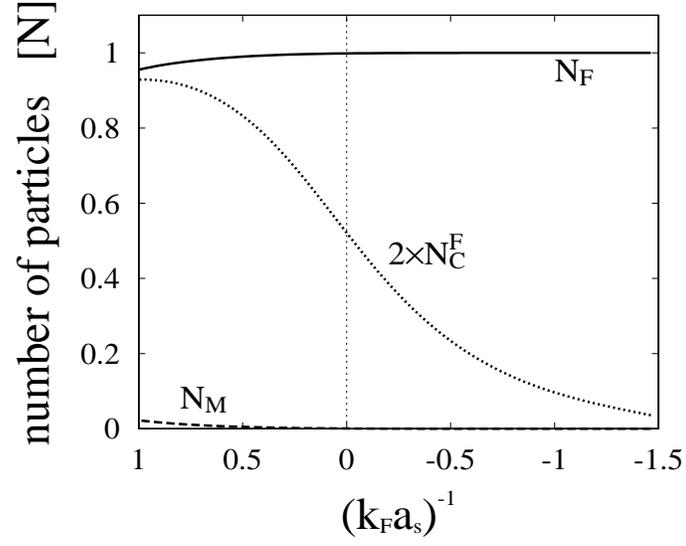}%
\caption{
Character of particles at $T=0$ in the crossover region in the case of a broad Feshbach resonance. $N_{\rm F}$ is the total number of Fermi atoms in the open channel, and $N_{\rm M}$ is the molecular condensate fraction (i.e., the number of Bose-condensed Feshbach molecules). $N_{\rm C}^F$ is the Cooper-pair condensate fraction, describing the BCS phase. 
\label{fig3} 
}
\end{figure}

\begin{figure}
\includegraphics[width=16cm,height=16cm]{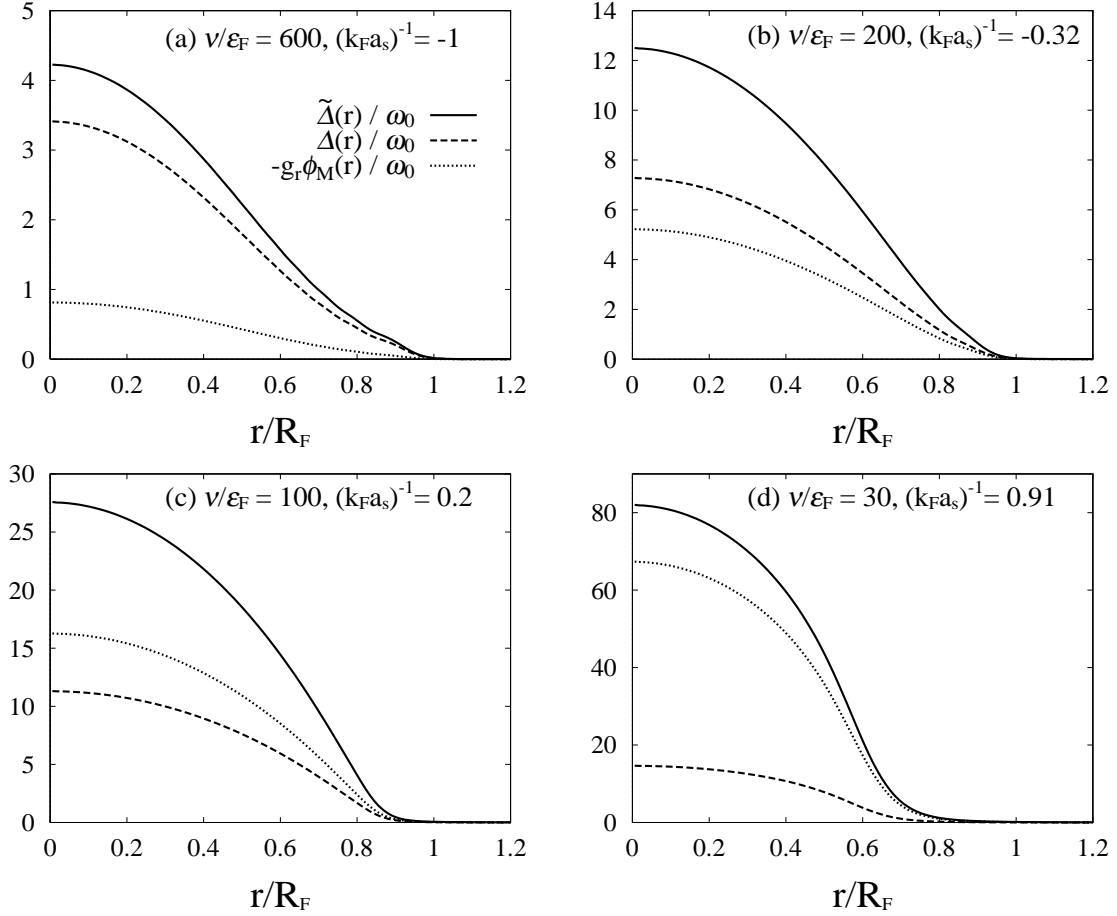}%
\caption{
Profile of the composite order parameter ${\tilde \Delta}(r)$ in the BCS-BEC crossover for a broad resonance. The separate contributions from the Cooper-pair component $\Delta(r)$ and the molecular BEC order parameter component $-g_{\rm r}\phi_M(r)$ are also shown. Compare with the analogous results for a narrow resonance shown in Fig. 7 of Ref.\cite{Ohashi}.
\label{fig4} 
}
\end{figure}

\begin{figure}
\includegraphics[width=10cm,height=13cm]{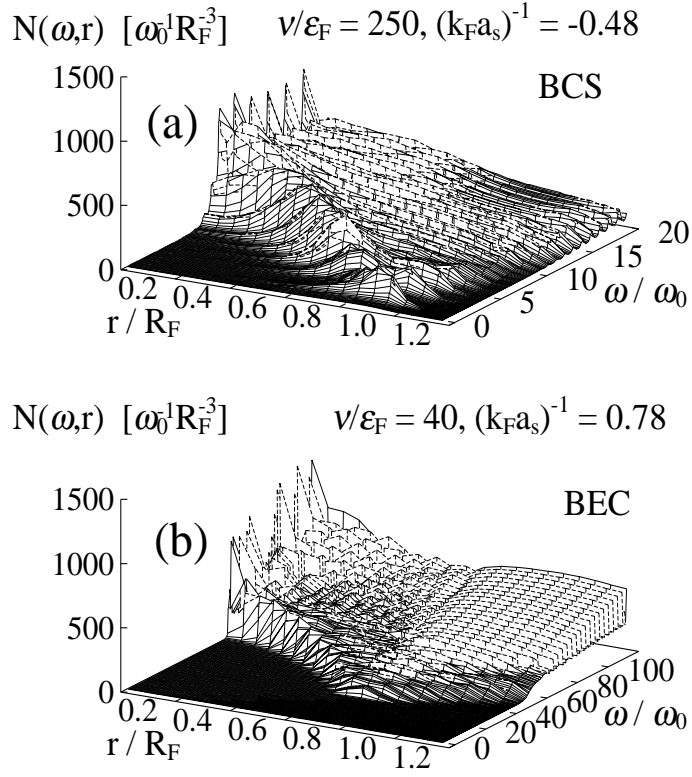}%
\caption{
Local single-particle density of states $N(\omega,r)$. (a) BCS region. (b) BEC region. We have introduced a small imaginary part $\Gamma=0.2\omega_0$ to the energies given by the BdG equations, to smooth out the results.
\label{fig5} 
}
\end{figure}

\begin{figure}
\includegraphics[width=10cm,height=15cm]{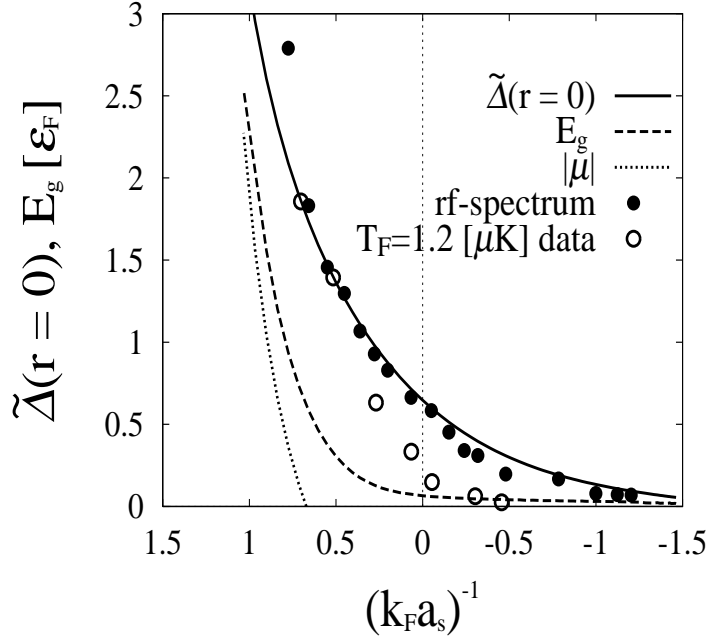}%
\caption{
Single-particle excitation gap $E_g$ in the BCS-like density of states $N(\omega,r)$, and the composite order parameter ${\tilde \Delta}(r=0)$ at the center of the trap. $E_g$ becomes large in the BEC region where the chemical potential $\mu$ is negative, and it approaches $|\mu|$ (dotted line) in this strong-coupling regime. The solid circles show the highest peak energies in the calculated rf-tunneling current spectra $I_{\rm F}(\omega)$ shown in Figs. 7 and 8. The open circles show the measured peak energy in the rf-tunneling spectrum in superfluid $^6$Li\cite{Chin}, for comparison. See Fig. 21 of Ref.\cite{Ohashi} for 
the analogous results for a narrow Feshbach resonance.
\label{fig6} 
}
\end{figure}

\begin{figure}
\includegraphics[width=10cm,height=16cm]{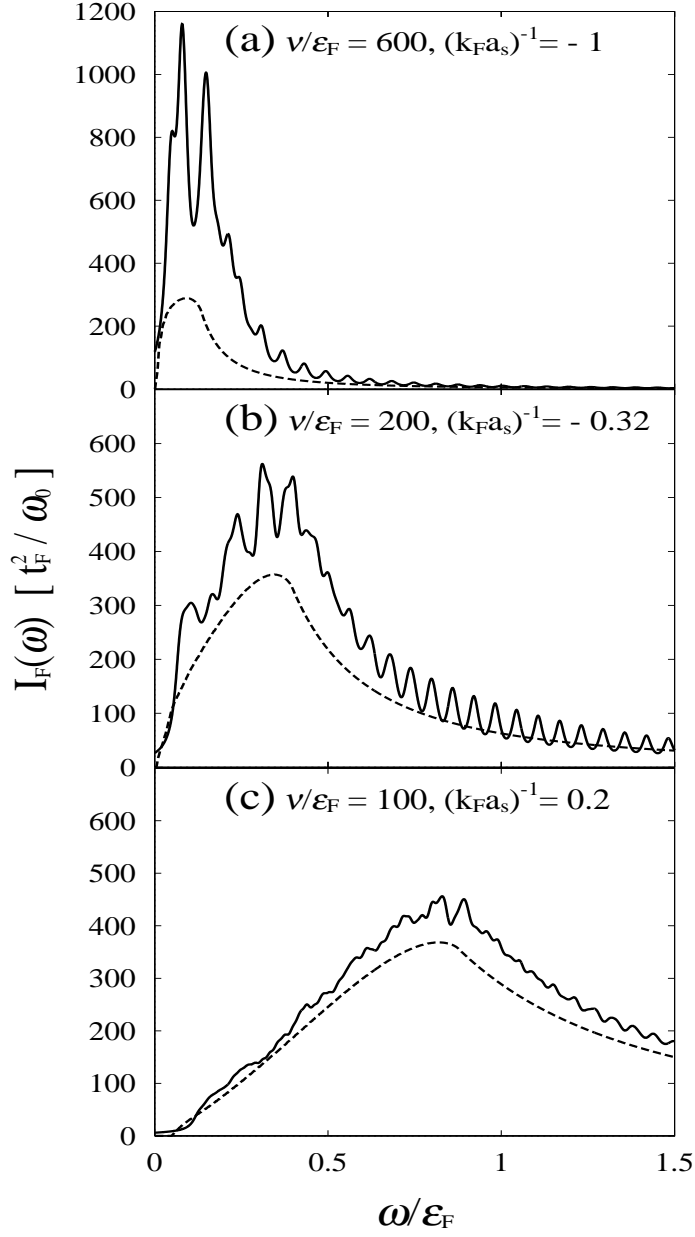}%
\caption{
Calculated rf-tunneling current spectrum in the BCS and crossover regions, based on the self-consistent solution of the BdG equations. The solid line shows the result using the microscopic expression for the rf-tunneling current spectrum, given by Eq. (7.10) of Ref.\cite{Ohashi}. The fine structure in the spectrum originates from the discrete BdG excitation energies in the harmonic trap. The dashed line is based on the LDA expression, given by Eq. (7.18) of Ref. \cite{Ohashi}. For clarity, we have introduced a small imaginary part $\Gamma=0.5\omega_0$ (damping) to the BdG eigenstates to smooth the results.
\label{fig7} 
}
\end{figure}

\begin{figure}
\includegraphics[width=10cm,height=13cm]{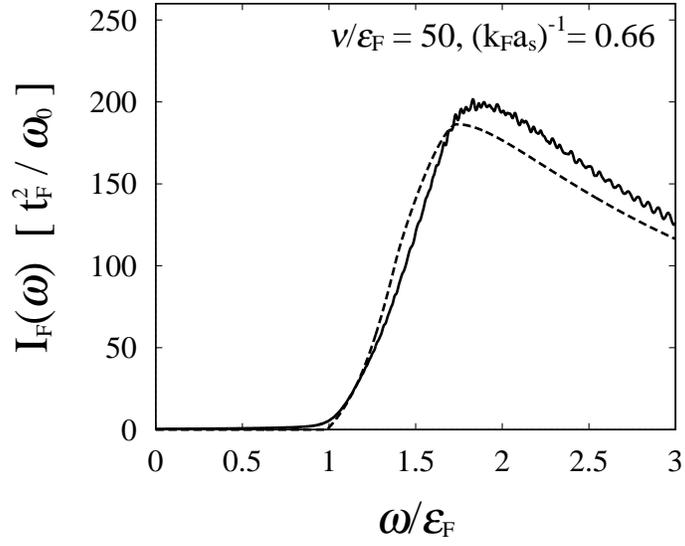}%
\caption{
Same plot as in Fig. 7, for $(k_{\rm F}a_s)^{-1}=0.66$ (the BEC regime).
\label{fig8} 
}
\end{figure}

\end{document}